# Two-dimensional Ferroelectric Ferromagnetic Half Semiconductor in VOF monolayer

Shaowen Xu[a], Fanhao Jia[a], Guodong Zhao[b,a], Wei Wu[a], and Wei Ren[a,†]

[a]*Physics Department, Shanghai Key Laboratory of High Temperature Superconductors, State Key Laboratory of Advanced Special Steel, International Centre of Quantum and Molecular Structures, Shanghai University, Shanghai 200444, China.*

[b]*School of Microelectronics, Fudan University, Shanghai, China*

†renwei@shu.edu.cn

Two-dimensional (2D) multiferroics have been casted great attention owing to their promising prospects for miniaturized electronic and memory devices. Here, we proposed a highly stable 2D multiferroic, VOF monolayer, which is an intrinsic ferromagnetic half semiconductor with large spin polarization ~2 $\mu_B$/V and a significant uniaxial magnetic anisotropy along *a*-axis (410 $\mu$eV/V atom). Meanwhile, it shows excellent ferroelectricity with a large spontaneous polarization 32.7 $\mu$C/cm$^2$ and a moderate energy barrier (~ 43 meV/atom) between two ferroelectric states, which can be ascribed to the Jahn-Teller distortion. Moreover, VOF monolayer harbors an ultra-large negative Poisson's ratio in the in-plane direction (~-0.34). The Curie temperature evaluated from the Monte Carlo simulations based on the Ising model is about 215 K, which can be enhanced room temperature under -4% compressive biaxial strain. The combination of ferromagnetism and ferroelectricity in the VOF monolayer could provide a promising platform for future study of multiferroic effects and next-generation multifunctional nanoelectronic device applications.

## Introduction

Two-dimensional (2D) multiferroics represent a class of unique materials[1] with two or more coexisting ferroic orders, such as ferromagnetism/antiferromagnetism, ferroelectricity, and ferroelasticity, which possess promising application prospects in miniaturized information storage and spintronic device[2-9]. The microscopic sources of ferroelectricity lead to two kinds of multiferroics namely type-I and type-II multiferroics[10-11]. The ferroelectricity and magnetic orderings in Type-I multiferroics have independent origins, which results in a weak magnetoelectric (ME) coupling with high transition temperature and large ferroelectricity polarization, e.g. 2D CuCrP$_2$S$_6$[12]. In the second group called type-II multiferroics, such as Hf$_2$VC$_2$F$_2$ monolayer[13] and VS$_2$ bilayer [14], the ferroelectricity polarization is induced by the magnetic ordering, thereby they often present a strong ME coupling with relatively low transition temperature and weak polarization. Unfortunately, due to the famous $d^0$ rules[15], most of 2D materials lack intrinsic polarization or ferromagnetic (FM) ordering, which limit their further applications in spintronics. Therefore, pursuing the multiferroic materials simultaneously possessing ferromagnetism and ferroelectricity still remains challenging hitherto [16-17].

Many researches are focused on antiferromagnetic (AFM) and ferroelasticity, for example, AgF$_2$ [18] and VF$_4$ [19] monolayers were proposed to have antiferromagnetic and ferroelastic coexistence; FeOOH monolayer was demonstrated to coexist three ferroics (antiferromagnetic, ferroelastic and ferroelectric) with significant electroelastic and magnetoelastic effects[20]; FeOX (X= Halogen element) [21] was found to have high Néel temperatures. All above interesting investigations inspired us to further explore the possibility of coexisting ferromagnetism and ferroelectricity in a single system, e.g., VOF monolayer in this work. Additionality, the auxetic materials have been extensively studied in the past decades, which expand laterally in the elastic range when stretched, while shrink in the transverse direction when compressed. Such negative Poisson's ratio (NPR) effect provides extraordinary mechanical performance and promising applications in electromechanical devices such as biomedicine and sports equipment [22-25]. However, the available auxetic 2D materials possess a relatively small NPR, for example, black phosphorene is predicted to have NPR -0.027[24]. Hence, it is highly desired to explore new 2D auxetic materials with large NPRs.

To our knowledge, 2D ferromagnetic ferroelectric multiferroics are still scarce with switchable ferro vectors simultaneously with high Cure Temperature ($T_c$), large

NPR, and strong anisotropy. Inspired by above mentioned literature, we propose a multiferroic VOF monolayer in this work. First of all, we demonstrate the stability from the dynamic, thermodynamic, and mechanical aspects. We find that the VOF monolayer is an intrinsic ferromagnetic half semiconductor with large spin polarization and strong magnetic anisotropy (410 $\mu$eV/V). Compared to other 2D auxetic materials, it also possesses a relatively large in-plane NPR (~-0.34). Additionality, the Curie temperature evaluated from the Monte Carlo simulations based on the Ising model is predicted to be ~ 215 K, which is close to room temperature when applying -4% biaxial compressive strain. Moreover, it is also demonstrated to be ferroelectric with the polarization ~32.7 $\mu$C/cm$^2$. These findings will enrich the family of 2D multiferroics with coexisting ferroelectric and ferromagnetic order, and pave a way for realizing the low dimensional nonvolatile logic and memory devices.

## Results and discussion

**Structural and Ferroelectric Polarization Properties**

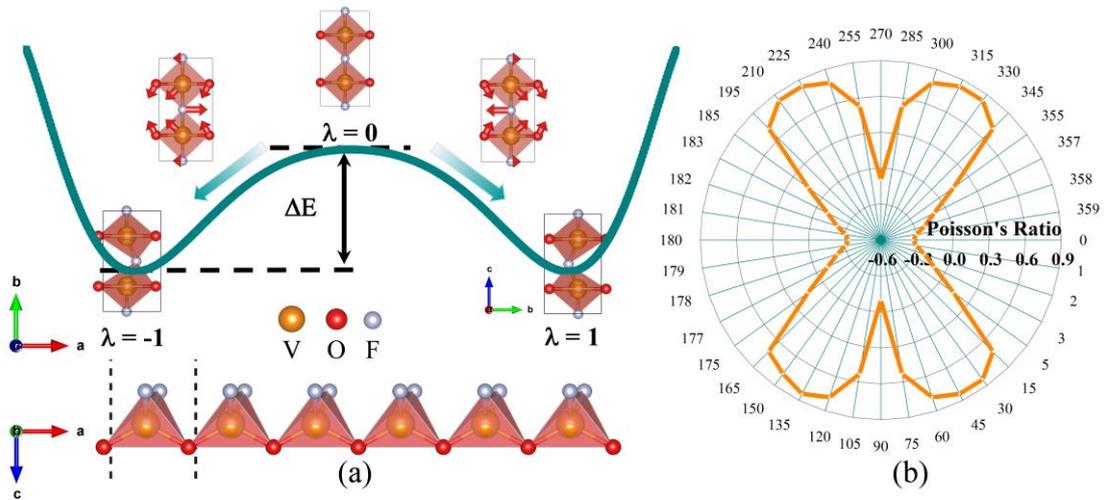

Figure 1. (a) Double well potential of ferroelectric reversal transition in the VOF monolayer, and side views of the VOF monolayer. The rectangles indicate the unit cells. λ=1, -1 and 0 describe the FE state-1, FE state-2 and centrosymmetric PE phase, respectively. (b) The orientation angle-dependent in-plane Poisson's ratio for VOF monolayer, where the θ = 0° represents the *a*-direction.

Bulk VOF is a member of transition metal oxyhalides, which are antiferromagnetic semiconductors with van der Waals (vdW) layered orthorhombic crystal structure (space group 59 *Pmmn*)[26-27]. The monolayer structure taken from the bulk experiences a significant distortion due to the symmetry broken along the c direction, of $C_{1h}$ point group, which is expected to present distinct properties. The top and side views of VOF monolayer are shown in Fig. 1(a).

To verify the stability of the VOF monolayer, we perform the phonon dispersion, molecular dynamic and mechanically simulations. From the perspective of lattice dynamics in Fig.S1(a), we find no appreciable imaginary mode in the phonon dispersion, indicating the dynamitic stability of VOF monolayer. The *ab initio* molecular dynamics (AIMD) simulations at 300 K were performed to check the thermodynamic stabilities as well. We observed in Fig. S1(b), the 2D planar networks and geometry shapes are well preserved, suggesting a robust thermal stability for the VOF monolayer. Moreover, the elastic constants of VOF monolayer meet the Born criteria: $C_{11}>0$, $C_{66}>0$ and $C_{11}*C_{22} > C_{12}*C_{12}$, which further confirm that it is mechanically stable.

The optimized structural parameters are displayed in Table I, which shows that the VOF monolayer has the lattice constants of $a = 3.18$ Å, $b = 6.13$ Å, the bond angles $\theta_{O-V-O} =122.55°$, $\theta_{F-V-F} = 98.67°$ and $\theta_{V-F-V}= 89.65°$. It is easy to see that these systems naturally present an out-of-plane symmetry breaking. It maintains a mirror symmetry plane which is perpendicular to the *b*-axis, so there is no polarization along the *b*-axis. Nevertheless, off-center V ions results in the in-plane polarizations along the *a*-axis in VOF monolayer. To confirm the occurrence of the spontaneous polarization, we adopt an adiabatic pathway from FE phase to PE phase by using an interpolation method. The energies as function of the amplitudes of the in-plane distortions are displayed in the Fig. 1(a), where an interpolation parameter λ is introduced to represent the displacements with respect to the high-symmetry structure. The calculated barrier energy is ~ 43 meV/atom, which is high enough to ensure the thermal stability of the spontaneous electric polarization at room temperature. Taken the thickness as 5 Å, the VOF monolayer possesses relatively large in-plane polarization calculated by using

Born effect charge[7] along the *a*-axis is ~32.7 $\mu C/cm^2$, which is comparable and even larger than pervious predicted 2D ferroelectrics, e.g. FeOOH monolayer with the thickness of 6 Å and a polarization ~ 13 $\mu C/cm^2$ [20]; Taking the thickness as 1 nm, the polarization was reported to be 30.2 $\mu C/cm^2$ for VOCl$_2$ monolayer[28].

Table I: Structural parameters of VOF monolayer. The lattice constant (Å), bond lengths and bond angles of VOF monolayers by PBE+U.

| Lattice Constant (Å) | Bond Lengths (Å) | Bond Angles (°) |
|---|---|---|
| a=3.18 | 2.0 (V-F$_1$) | 98.67(F-V-F) |
| b=6.13 | 2.1 (V-F$_2$) | 89.65(V-F-V) |
|  | 1.86 (V-O) | 122.6(V-O-V) |

To better understand the origin of ferroelectricity, we plot the *d* orbital resolved density of states of ferroelectric (FE) and paraelectric (PE) phases calculated by HSE06 functional in Fig.S2. We can easily find the band gap of FE phase is larger than PE phase. For the case of centrosymmetric PE phase, the energy gap between $d_{xy}$ and $d_{xz}+d_{yz}$ orbitals in the proximity of the Fermi level is ~ 0.91 eV. In the FE state, the V ions are displaced along the *a*-axis, resulting the further splitting of $d_{xy}$ and $d_{xz}+d_{yz}$ orbitals, and widening energy gap (~1.11 eV). Consequently, the total energy of the FE state is lower than that of the PE state. The ferroelectric displacement of V atoms could be further understood by the Jahn-Teller distortion mechanism[29].

The calculated elastic constants are listed in Table SI. The in-plane Poisson's ratio of orthogonal symmetry monolayers as a function of $\theta$ is shown in Fig. 1(b) and calculated by the following equation[19, 24]:

$$v(\theta) = \frac{C_{12}\cos^4\theta - B\cos^2\theta\sin^2\theta + C_{12}\sin^4\theta}{C_{22}\cos^4\theta + A\cos^2\theta\sin^2\theta + C_{11}\sin^4\theta} \qquad (1)$$

Here, the A = $(C_{11}C_{22} - C_{12}^2)/C_{66} - 2C_{12}$ and B = $C_{11} + C_{22} - (C_{11}C_{22} - C_{12}^2)/C_{66}$, and $\theta$ is the angle of the direction with respect to the *a*-axis. We find that the Young's modulus for the VOF monolayer varies from 25 N/m to 87 N/m, suggesting the significant mechanical anisotropy of the VOF monolayer. In addition, the in-plane Poisson's ratio presented in Fig.1 (b) exhibits anisotropy along the $x$ (~ -0.34) and $y$ (~-0.09) directions. Notably, the largest negative and positive Poisson's ratios can reach -0.34 and 0.84 along the $x$ direction and the diagonal direction, respectively. The NPR is much larger than the recently reported 2D auxetic materials, such as δ-phosphorene (-0.267) [24] and monolayer black phosphorus with a puckered structure (-0.027)[30]. In the directions away from $x$ and $y$ axes, VOF monolayer presents a normal positive Poisson's ratio in general.

**The Electronic and Ferromagnetic Properties**

The band structure and the spin-resolved density of states (DOS) calculated by PBE+$U$ (U= 3 eV) are displayed in Fig.2(a) and Fig.2(b), showing the VOF monolayer is a direct semiconductor with the band gap around ~0.42 eV for PBE+$U$ method (~1.7 eV for HSE06 method). The valence band maximum (VBM) and conduction band minimum (CBM) are both located at Γ (0,0) point. Obviously, we can observe the band dispersions along the Γ-X (reciprocal lattice equivalent of the crystallographic $a$-axis) are steeper than Γ-Y directions near Fermi level, which indicate the electronic anisotropy. From Fig.2(b), we find both the VBM and CBM are predominantly contributed by the V atoms. In more detail, the VBM is majorly originated from V-$d_{xy}$ orbitals with a slight hybridization of O-2$p_y$ states, and the CBM is mainly derived from empty V-$d_{xz}$ orbitals. Remarkably, the VBM and CBM are fully spin-polarized and show the same spin channel, which can be categorised to the class of 2D half semiconductors [31]. A large spin exchange splitting Δ = 1.3 eV is found in the conduction band, making possible to achieve half-metallic characteristic in VOF monolayer. A potential strategy of realizing the half-metal might be applying a negative gate voltage in experiment to inject electrons into the system and produce an electron doping effect[32]. Hence, we further investigate the carrier-doped electronic properties, demonstrating there exists a semiconductor to half-metal phase transition under carrier doping [see Fig.S3]. In particular, the hole-doped VOF monolayer presents a complete spin-up polarization, while the electron-doped is first characterized by fully spin-up polarized and then metal properties. Due to the large spin exchange splitting, the half-metallicity can be preserved under a high hole doping density, which is feasible in experiment by using an ionic liquid as a gate dielectric[32-33]. All above mentioned results give a potential approach to manipulate the carrier's spin-polarization in VOF monolayer and further simulate the fabrication of the spin-electronic device such as dual-channel field effect spin-filter and spin-valve[34].

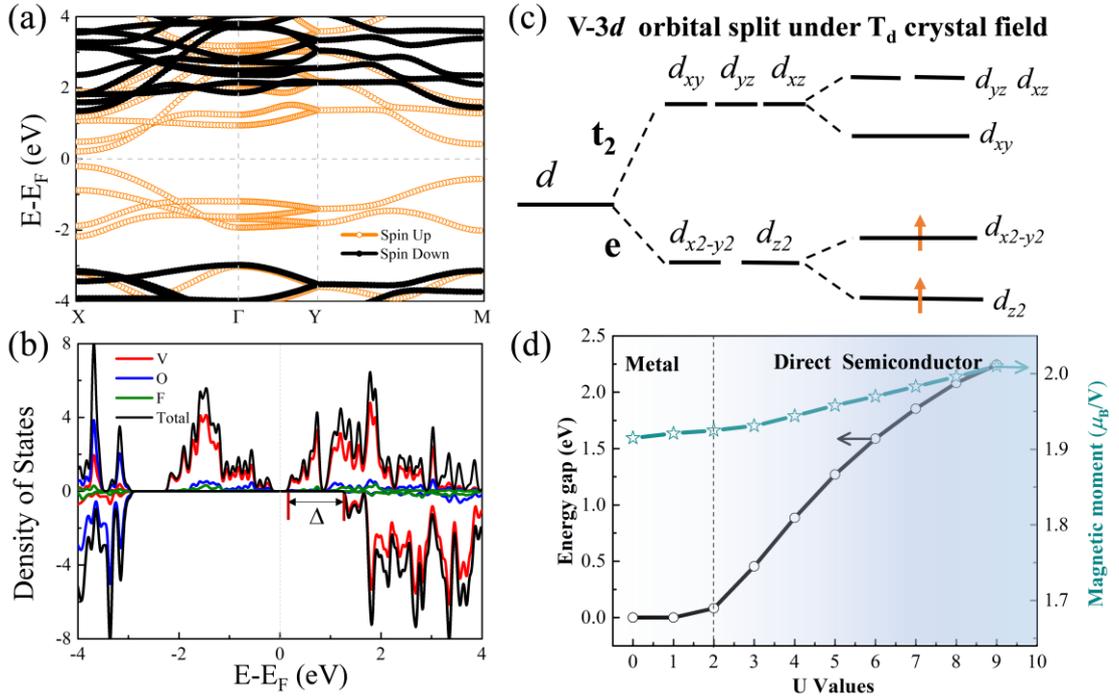

Figure 2. The (a) band structure and (b) partial density of states of VOF monolayer. (c) The schematical illustration of the splitting of the V-3$d$ states under a tetrahedral crystal field. (d) The energy gap and magnetic moment of VOF monolayer as functions of U values.

We determine the magnetic ground state by building 2×1 supercell with four different spin arrangements namely FM, AFM-I, AFM-II and AFM-III, which are illustrated in Fig.S4. The results show that the FM state is the most energetically favourable in all considered structures. The band gaps of VOF monolayer are sensitive to the magnetic ordering, as listed in Table SII. From Fig.S5(a), most of the spin-polarized electrons are locate around the V ions, while O atoms carry only tiny local magnetic moments. The magnetic moment is 1.93 $\mu_B$/V atoms, which can be explained by the electronic configuration of a V atom $[Ar]3d^34s^2$. Due to the bonding to neighbouring O and F atoms, the V atom loses $3e^-$ and becomes $V^{3+}$ with an electronic configuration of $[Ar] 3d^2$. According to the Hund's rule, the remaining $2e^-$ on V occupy different $d$ orbitals so that the magnetic moment of V should be $2\mu_B$, close to that obtained from DFT calculations.

Based on the crystal field splitting theory (see Fig.2(c)), we can establish the magnetic exchange paths along the $a$-axis of V-O-V and V-V chains. The O-mediated super-exchange interaction favors a ferromagnetic coupling between the adjacent V atoms ($d_{xy}$-$p_{x/y}$-$d_{xy}$), while the direct-exchange interaction gives rise to an antiferromagnetic coupling between the adjacent V atoms. According to the Goodenough–Kanamori–Anderson rules[35-36], FM coupling is favored when the superexchange interaction chains are around 90°, otherwise the AFM coupling is preferred. The bond-angle of V-O-V here (Table I) is close to 90°, which suggests V-$d$ orbitals are nearly orthogonal to the O-$p$ orbitals. We further reveal the real-space wavefunction squared of the highest valence band at the gamma point in the Fig.S5(b), which shows the coexistence V: $d_{yz}$ and O: $p_y$ orbitals and the absence of F orbitals. Therefore, we attribute the stable FM ordering to the competition between the super-exchange and direct-exchange interactions based on the Goodenough-Kanamori-Anderson rules[37].

In addition, we checked the effect of Hubbard-$U$ corrections in Fig.2(d). Both electronic and magnetic properties are sensitive to the selected $U$ values. The band gap increases from the value of 0 eV to 2.25 eV (when $U_{eff}$ = 9 eV). The VOF monolayer turns out to be metallic when $U$ < 2 eV, while behaves as a semiconductor for larger $U$ over 2 eV, accompanying a metal-insulator transition. It is worth mentioning that different $U$ values do not change the ferromagnetic and ferroelectric ground state, it only modifies the magnetic moment by about 0.1 $\mu_B$ from $U$=0 to 9 eV. Thus, we conclude that the basic properties are not altered by the choice of Hubbard $U$ in DFT calculation.

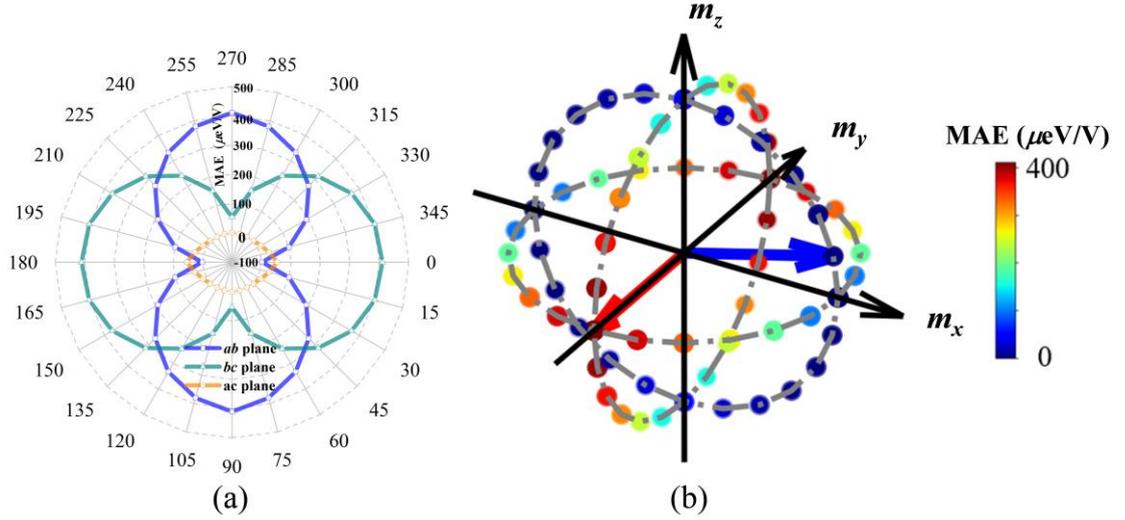

Figure 3. The angular dependence of the MAEs of VOF monolayer with the direction of magnetization lying on three different projected planes (a) and the three-dimensional illustration (b).

Magnetic anisotropy is important to utilize the long-range ferromagnetic ordering[38], hence, we consider the spin-orbit coupling (SOC) to evaluate the angle dependent magnetic anisotropy energy (MAE) of VOF monolayer as displayed in Fig.3. It clearly shows that the MAE strongly depends on the direction of magnetization in the *ab* and *bc* planes, while it is much less isotropic in the *ac* plane. The easy magnetization axis is in the *ab* plane and has an angle with *b*- direction of 75°, which makes the VOF monolayer belong to 2D Ising ferromagnetic system[39]. The MAE reaches a maximum value of 410 $\mu$eV per V, which is comparable to that of the CrI$_3$ (685 $\mu$eV/Cr)[40], GdI$_2$ (553 $\mu$eV/Gd)[9], and larger than MnAs monolayer (281 $\mu$eV/Mn)[41] and CrP (217 $\mu$eV/Cr)[42].

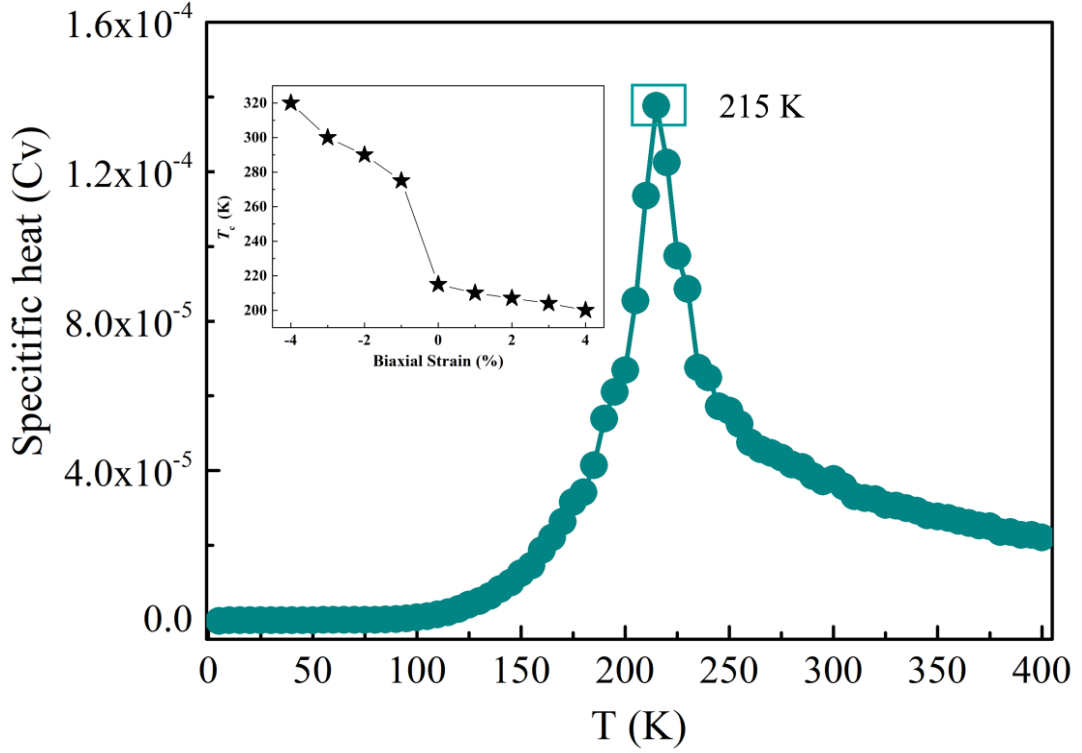

Figure 4. (a) The specific heat capacity (Cv) as a function of temperature for the VOF monolayer. The schematic representation of the exchange interaction is presented in the Figure S4. The inset is the Tc under different biaxial strains of VOF monolayer.

Generally, the Curie temperature $T_c$ is positively correlated with the MAE. We further perform the Monte Carlo (MC) simulations based on the Ising model to estimate the $T_c$, which was widely used in previous studies. For simplification, we only considered the nearest-neighbor (NN) and next-nearest-neighbor (NNN) exchange interactions. The NN exchange coupling parameters along the $a$- and $b$-directions are respectively $J_a$ and $J_b$, the NNN exchange coupling parameter is $J_{ab}$, as indicated in the insets of Fig. S3. The Hamiltonian is written as:

$$\mathbf{H} = -2J_a \sum_{\langle ij \rangle} \mathbf{S}_i \cdot \mathbf{S}_j - 2J_b \sum_{\langle\langle ij \rangle\rangle} \mathbf{S}_i \cdot \mathbf{S}_j - 2J_{ab} \sum_{\langle\langle\langle ij \rangle\rangle\rangle} \mathbf{S}_i \cdot \mathbf{S}_j$$

The total energies of four magnetic configurations are expressed below:

$E_{FM}/4 = E_0 - 2S^2(J_a + J_b + 2J_{ab})$     (2)

$E_{AFM-I}/4 = E_0 - 2S^2(J_a - J_b - 2J_{ab})$     (3)

$E_{AFM-II}/4 = E_0 - 2S^2(-J_a - J_b + 2J_{ab})$     (4)

$$E_{\text{AFM-III}}/4 = E_0 - 2S^2(-J_a + J_b - 2J_{ab}) \quad (5)$$

Here, $E_0$ is the energy of the nonmagnetic part and $S$ is the spin magnetic moment of VOF monolayer. By solving these equations, we obtain the three exchange coupling parameters $J_a$, $J_b$ and $J_{ab}$, which are implemented in the MC simulations. We adopt 50 × 50 ×1 square supercell (2500 local magnetic moments) in the simulations. The specific heat $C_v$ as a function of temperature is displayed in Fig.4, which has a peak around $T_c$ = 215 K, indicating the ferromagnetic-paramagnetic phase transition. Although the $T_c$ is slightly lower than room temperature, it might be effectively tuned to be close to room temperature by biaxial compressive strains. The $T_c$ of VOF monolayer under different biaxial strain is displayed in the inset of Fig.4. The $T_c$ of the VOF monolayer is enhanced to ~320K by applying -4% biaxial compressive strain, which indicates the intrinsic magnetic field of VOF monolayers is likely to survive at room temperature. We also compared the energy difference between the ferromagnetic and antiferromagnetic states, found that the ferromagnetic is always energetically advantageous under external strains and electric fields, suggesting the FM VOF monolayer is robust against external condition. Furthermore, there are 2D transition metal oxyhalides and nitride monolayers that have been predicted to be ferromagnetic materials with a high Curie temperature, such as CrOCl[43], TcOBr [44] and MnNX (X=F, Cl, Br and I)[45]. Another interesting question to consider is about the coupling between ferroelectric and ferromagnetic order. For VOF monolayer system, the easy axis of magnetization is almost perpendicular to the ferroelectric polarization direction, which provides possible direct electric-field control of the magnetization.

**Conclusions**

In summary, from the first-principles we demonstrate that ferroelectricity and ferromagnetism simultaneously exist in VOF monolayer. The unique crystal structure leads to in-plane spontaneous electric polarization along the *a*-direction and moderate activation energy barrier. V ions with spin magnetic moment ~2 $\mu_B$/V are aligned in a ferromagnetic order. The Curie temperature was predicted to be about 215 K from the Monte Carlo simulations based on the Ising model. The easy magnetization axis is almost perpendicular to the direction of electric polarization. Our findings open a new route for exploring 2D multiferroic materials and building spintronic devices.

**Computational Method**

The first-principles calculations are carried out within the generalized gradient approximation (GGA) frame proposed by Perdew, Burke, and Ernzerhof (PBE) [46], which is implemented in the Vienna *ab initio* simulation package (VASP) [47]. We chose the energy cutoff to be 600 eV, and an additional Dudarev's effective Hubbard $U_{eff}$ = 3.0 eV for V-3d orbitals to deal with the self-interaction error[48]. The $\Gamma$ centered k-grids were adopted to be 6×6×1. The atomic positions were fully relaxed until the maximal force on each atom was less than $10^{-4}$ eV/Å. To solve the well-known problem of underestimated band gap of DFT exchange-correlation functionals, the screened hybrid functional HSE06 [49] was applied to calculate the band gaps. We adopted a vacuum layer thicker than 20 Å to avoid the interaction between periodic images. The spin-orbit coupling (SOC) was also considered to take the orientation dependence of spin and orbital magnetic moments into account. The phonon dispersion was calculated by using density-functional perturbation theory as implemented in the PHONOPY package [50-51], which is also valid in 2D systems[7, 52]. The calculated polarization values of 2D materials strongly depended on the thickness of slab we chose[7]. In our calculations, we added two additional 1 Å along the *c*-axis on two sides of VOF monolayer for the calculation of volume $\Omega$. The Born effective tensor is defined as: $Z^*_{i,j} = \frac{\Omega}{e}\frac{\partial p_i}{\partial u_j}$, in which *e* is the charge of electron, *p* is polarization, and *u* is the atomic displacement from its high-symmetry nonpolar position.

**Conflicts of interest**

The authors declare no competing financial interest.

**Acknowledgements**

This work was supported by the National Natural Science Foundation of China (51861145315, 11929401, 12074241), the Independent Research and Development

Project of State Key Laboratory of Advanced Special Steel, Shanghai Key Laboratory of Advanced Ferrometallurgy, Shanghai University (SKLASS 2020-Z07), the Science and Technology Commission of Shanghai Municipality (19DZ2270200, 19010500500, 20501130600), and High Performance Computing Center, Shanghai University. F.J. is grateful for the support from the China Scholarship Council (CSC).